\documentclass[conference,romanappendices,10pt]{IEEEtran}	

\IEEEoverridecommandlockouts

\usepackage{algorithm,algorithmic,amsmath,amssymb,amsthm,array,bibentry,cite,color,comment,enumerate}
\usepackage{enumitem,eurosym,float,graphicx,lettrine,mathrsfs,multicol,multirow,nomencl,pict2e,psfrag}
\usepackage[absolute]{textpos}
\usepackage[USenglish]{babel}

\renewcommand\thealgorithm{A\arabic{algorithm}}
{\newtheorem{theorem}{Theorem}}
 
\begin{document}
\title{Optimizing Access Mechanisms for QoS Provisioning in Hardware Constrained Dynamic Spectrum Access}

\author{
\IEEEauthorblockN{Spyridon~Vassilaras and George~C.~Alexandropoulos}
\IEEEauthorblockA{Mathematical and Algorithmic Sciences Lab, France Research Center, Huawei Technologies Co$.$ Ltd$.$, \\20 Quai du Point du Jour, 92100 Boulogne-Billancourt, France}
emails: \{spyridon.vassilaras, george.alexandropoulos\}@huawei.com}

\maketitle
\begin{abstract}
One of the major challenges in Dynamic Spectrum Access (DSA) systems is to guarantee a required level of Quality of Service (QoS) to secondary users of the spectrum. In this paper, we propose efficient algorithms for deriving optimal policies for the sensing / transmitting trade-off in hardware-constrained DSA systems. Unlike previous approaches which seek to maximize mean data rate for the secondary users, the proposed algorithms derive policies which minimize the probability of excessive queuing delays. Large Deviations (LD) asymptotics are used to approximate the probability of interest and policies maximizing the associated LD exponent are proposed. Although dynamic programming is not able to identify the optimal policy in this case, much more efficient algorithms than exhaustive search are proposed. These algorithms are based on specific properties of the optimal policy which are described and proven in this paper. 
\end{abstract}


\section{Introduction}\label{sec:Introduction}
The concept of Cognitive Radio (CR) has been researched and developed for many years but its commercial success has been far below initial expectations. New paradigms such as Licensed Shared Access (LSA), Authorized Shared Access (ASA), and Licensed Assisted Access (LAA) have been introduced as an evolution of the original CR schemes. These more conservative approaches to spectrum sharing sacrifice the benefits of Dynamic Spectrum Access (DSA) based on sensing, in favor of static sharing based on database information on spectrum usage. This approach was adopted in order to ease the concerns of Primary Users (PUs) of the spectrum for adverse effects on their networks and the concerns of Secondary Users (SUs) of low resource availability, and the resulting inability to guarantee a certain level of Quality of Service (QoS) to their customers. In spite of these developments, the CR community has not given up the effort to develop advanced DSA techniques that will be able to address the above concerns, while exploiting unused spectrum in a more efficient manner. It is hoped that such technical evolutions will make their way to standards and regulation in future releases of LSA / ASA and LAA related standards. 

Even since the need to provide some sort of QoS guarantees to SUs was recognized, there has been significant effort in identifying, estimating, and improving a number of SU-QoS parameters in DSA systems. In \cite{Su_2008}, a cross-layer based opportunistic multi-channel medium access control (MAC) protocol for wireless ad hoc networks has been introduced, which integrates the spectrum sensing at the physical (PHY) layer with the packet scheduling at the MAC layer. The authors in \cite{C:Kartheek_2012} present generalized opportunistic splitting algorithms using stochastic approximation for allocating the available spectrum to the CR users. In \cite{C:Vassaki_2011}, a QoS-driven power allocation scheme was proposed that takes into consideration the interference of PUs to the secondary user in order to accomplish the optimal allocation. Focusing on hybrid satellite/terrestrial networks, \cite{J:Vassaki_2013} presented a power allocation algorithm that optimizes the effective capacity of the terrestrial link for given QoS requirements while guaranteeing a specified outage probability for the satellite link. In \cite{C:Ververidis_2012}, the authors show through extensive simulations that given some information on the expected level of activity in the video, the duty cycle of the PU alone can yield good predictors for the expected quality of experience of SUs. Considering multi-antenna scenarios, statistical optimization techniques are applied in \cite{J:Imrana_2013} to assess the performance of the QoS in CR systems, where each user has different demands. In \cite{C:Ishibashi_2008}, it was shown that virtual wireless networks can be created, utilizing only the residual wasted bandwidth of the primary service providers. These virtual networks are able to support large volumes of users, while still ensuring that QoS reliability requirements, such as blocking and dropping guarantees, are achieved. Considering the time-varying arrival and service process, in \cite{C:Zhang_2009} the authors obtained analytically the QoS metrics, such as the approximated delay-violation probability and mean time delay. Very recently, the main characteristics of the notion of LSA / ASA have been proposed \cite{C:Mueck_2014} and relevant architectures as well as techniques have started to being introduced (e$.$g$.$ \cite{C:Mueck_2014,C:Taramas_2014}).

In this paper, we study the problem of providing QoS guarantees to the SUs in a DSA system and propose efficient algorithms to optimize certain QoS-related parameters in such a setting. In particular, we are addressing the well-known problem of optimizing the sensing / transmitting trade-off in a system with hardware constrains, as identified in \cite{Jia_2008}. These practical hardware constrains impose the limitation that a SU cannot: \textit{a}) sense and transmit at the same time; and \textit{b}) sense the whole range of spectrum at the same time. In \cite{Jia_2008}, sensing / transmitting policies that optimize the average SU throughput have been presented. Hereinafter, we propose a policy that minimizes a Large Deviations (LD) parameter which determines the probability that the queue size (and thus the associated queuing delay) will exceed a certain threshold.  

\section{Problem Formulation}\label{sec:Problem}  
We are using the same sensing and spectrum access model as in \cite{Jia_2008}. The SU performs wideband spectrum sensing by breaking the total bandwidth into $W$ channels, denoted as ${\rm ch}_1,{\rm ch}_2,\ldots,{\rm ch}_W$, and performing sensing in each channel sequentially. This sequential spectrum sensing follows from the common assumption that a CR is capable to sense limited bandwidth of spectrum during a certain amount of time $\tau$, which we call a sensing time slot from now on. For this reason the SU can devote a certain number of sensing time slots to sensing, and then use the discovered idle channels for transmitting its information bearing signals (see \cite[Fig$.$~2]{Jia_2008}). In order to avoid large transmission overlaps with PUs that start their transmissions after sensing has been performed, each SU transmission lasts for a short duration after which the sensing procedure is repeated. As a result, in each sensing / transmission period $T$, the longer sensing lasts (more channels can thus be sensed) the shorter the transmission time will be (if $k$ sensing time slots are used, the transmission time is $T-k\tau$). Hence, there is a trade-off in deciding when to stop sensing additional channels and start transmitting in the idle channels discovered so far. We assume that this decision can be taken dynamically at each time slot based on a predetermined optimal stopping policy, the number of idle channels discovered so far, and the knowledge of the (possibly joint) probability distribution of finding channels ${\rm ch}_1,{\rm ch}_2,\ldots,{\rm ch}_W$ idle. In \cite{Jia_2008}, the proposed optimal stopping policy aims at maximizing the expected throughput of SUs. 
\begin{figure}[t!] 
\centering
\includegraphics[width=3.2in]{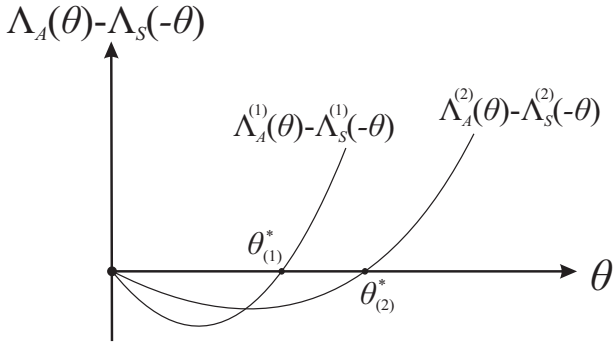} 
\caption{Graphical representation of two $\Lambda_A(\theta) - \Lambda_S(-\theta)$ functions and their associated roots.}
\label{Fig:Figure_1}
\end{figure}

It is well known that, when considering QoS parameters such as queue lengths and queuing delays, maximizing the mean service rate is not necessarily the optimum approach to optimizing these parameters (see, e$.$g$.$, \cite{J:Vassilaras_2009}). The variance and higher order moments of the service process affect queuing behavior as well. In this paper, we consider the probability of excessive queuing delay (over a given threshold), which is a common QoS metric, and show how to efficiently derive the optimal stopping rule for minimizing this probability (or more precisely, minimizing the exponent of the LD approximation of this probability). The following discrete time queuing model is adopted: time is divided into time slots equal to one sensing / transmitting period $T$. The amount of packets arriving to the queue during a time slot is denoted by $A(t)$ and modeled as a Markov chain with $M$ states, while service $S(t)$ is an independent and identically distributed (i$.$i$.$d$.$) process the probability density function of which can be affected by the choice of the stopping rule policy. We assume that $T=K\tau$ with $K\geq W$, and that the number of transmitted packets in a sensing / transmitting period $T$ is equal to $s(K-k)c$, where $s$ is the number of discovered idle channels and $c$ represents the number of packets that can be transmitted in one channel in time $\tau$. We model the buffer at the SU transmitter as an infinite queue (see \cite[Fig$.$~1]{J:Vassilaras_2009}) which operates in a first-in-first-out (FIFO) mode. We seek to minimize the steady state probability of the excessive queuing delay, defined as 
\begin{equation}\label{Eq:Probability}
\mathcal{P}_d\triangleq\mathbb{P}\{\text{delay}>D_{\rm max}\}
\end{equation}
where $D_{\rm max}$ is a maximum tolerable delay (a QoS parameter). We assume that not only the queue can be made stable if an appropriate sensing / transmitting policy is applied, but the resulting $\mathcal{P}_d$ can be made very small in order to meet the desired QoS level.

The probability $\mathcal{P}_d$ can be approximated using LD theory, as explained in \cite{J:Vassilaras_2009},  according to the following steps:
\begin{enumerate}
\item Find the limiting log-moment generating functions $\Lambda_A(\theta)$ and $\Lambda_S(\theta)$ of the arrival $A(t)$ and service $S(t)$ processes, respectively. Then, determine the positive solution $\theta^*$ of the equation $\Lambda_A(\theta) - \Lambda_S(-\theta) = 0$, and denote $\delta\triangleq\Lambda_A(\theta^*) = \Lambda_S(-\theta^*)$.
\item For any desired delay bound $D_{\rm max}$, the delay bound violation probability can be then approximated as $\mathcal{P}_d \cong \exp(-\theta^*\delta D_{\rm max})$. This approximation is accurate for small $\mathcal{P}_d$ (e$.$g$.$, in the order of $10^{-3}$ or smaller) as it is derived by an asymptotic expression (i$.$e$.$, for $D_{\rm max}\rightarrow\infty$). 
\end{enumerate}
It is well known that the function $\Lambda_A(\theta) - \Lambda_S(-\theta)$ is convex, equal to zero at the origin and has at most one positive solution (denoted by $\theta^*$) as shown in Fig$.$~\ref{Fig:Figure_1} (if it has no positive solution we say that $\theta^*=+\infty$). It is noted that $\Lambda_S(\theta)$ can be controlled by the selected control policy. Also, note that $\Lambda_A(\theta)$ is an increasing function of $\theta$ and, therefore, the policy which maximizes $\theta^*$ will also maximize $\delta$, and thus minimize the LD approximation of $\mathcal{P}_d$. We, henceforth, turn our attention into determining the policy maximizing $\theta^*$.

\begin{figure}[t!] 
\centering
\includegraphics[width=3.2in]{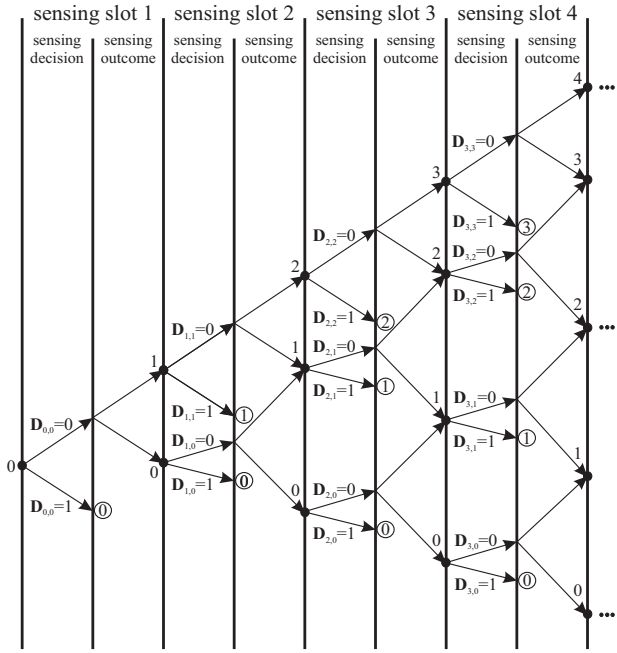} 
\caption{The decision ``tree'' (in fact the decision graph) of any sensing / transmitting policy.}
\label{Fig:Figure_2}
\end{figure}

Any sensing / transmitting policy can be represented by a $W\times W$ lower diagonal matrix $\mathbf{D}$ with elements $[\mathbf{D}]_{k,s}$ ($k=0,1,\ldots,W-1 \text{ and } s=0,1,\ldots,k$) denoting the decision when $s$ idle channels have been discovered after sensing $k$ channels. When $[\mathbf{D}]_{k,s}=0$, the decision is to continuing sensing whereas, $[\mathbf{D}]_{k,s}=1$ represents the decision to stop sensing and start transmission in the already discovered idle channels. The policy should be determined offline based on the knowledge of the arrival process and the probabilities of finding a channel idle. In this paper, we will only consider the case where the probability of finding a sensed channel idle is constant and independent of the sensing outcomes in other channels, and the outcome of sensing the same channel in previous sensing / transmission periods. Let us denote this probability by $p_{idle}$. Cases with unequal and / or dependent probabilities will be considered in future work. Although using matrix $\mathbf{D}$ is a compact way to represent a sensing / transmitting policy, a decision ``tree'' visualization, as the one shown in Fig$.$~\ref{Fig:Figure_2}, is a good way of understanding a policy and will also be useful in proving related results. Note that black filled circles in this figure represent states when a sensing / transmitting decision must be made whereas, circles with a number inside denote ``leafs'' of the ``tree'', i$.$e$.$, terminating states where sensing stops and transmission begins (the number in the circles indicates how many idle channels have been discovered so far). As some decision states can be reached through many different paths, this is not a real tree but rather a decision graph. It is possible to draw a real decision tree by splitting the said states to a number of different states, however it is assumed that the policy must make a decision based on the number of idle channels discovered so far and not on which particular channels have been found idle. It is easy to see that this restriction is not resulting to worse policies, as knowing the exact order of idle channel discoveries carries no information on subsequent channel sensing. In the following we will use the terms ``decision graph'' to refer to the graph in Fig$.$~\ref{Fig:Figure_2} and ``decision tree'' to refer to its transformation to a tree by an appropriate splitting of states. We will also use the term ``relaxed policy'' to denote a policy that can make independent decisions at different nodes of the decision tree that are the same nodes in the decision graph. In a slight abuse of notation, both a policy and its corresponding matrix will be denoted by a capital bold letter (e$.$g$.$, $\mathbf{D}$). A relaxed policy will also be represented by a capital bold letter, although it cannot be represented by a matrix in the same way as a (non-relaxed) policy. 

\section{Maximizing the LD Exponent}\label{sec:LD_Exponent}
An exhaustive search approach over all possible values of the policy matrix $\mathbf{D}$ is guaranteed to discover the optimum policy $\mathbf{D}^*$ which maximizes the LD exponent $\theta^*$. However, the exhaustive approach has a high computational cost and becomes quickly impractical as the size of the matrix $\mathbf{D}$ grows (i$.$e$.$, as $W$ increases). In \cite{Jia_2008}, a Dynamic Programming (DP) approach was shown to be able to find the policy which maximizes the average throughput in the same scenario in an efficient way.

In order to maximize $\theta^*$ we have experimented with the following approaches:
\begin{itemize}
\item Using a throughput maximizing policy can result to a $\theta^*$ (and an associated $\mathcal{P}_d$) that is sometimes identical or very close to $\theta^*$. However, optimality is not guaranteed.
\item Using a DP approach to find a $\theta^*$ maximizing policy often results into policies that are not optimal, and in some cases are performing even worse than a throughput optimizing policy. 
\end{itemize}
To overcome the limitations with the aforementioned approaches, we have devised two more efficient approaches in searching the control policies space for the optimal policy. In order to justify these two algorithms, we will first establish two theorems describing the form of the optimal policy matrix.  
In the following, we assume that the log moment generating function of the arrival process $\Lambda_A(\theta)$ is known, i.e., it is possible to calculate or estimate $\Lambda_A(\theta)$ for any given $\theta \geq 0$. The log moment generating function of the service process $\Lambda_S(\theta)$ is easy to calculate for a given $\mathbf{D}$ and $p_{idle}$: for any leaf in the decision tree we can calculate the resulting service rate $r_i = s(K-k)c$ and the probability $p_i = p_{idle}^{k-s}(1-p_{idle})^{s}$ of terminating at this leaf, and then calculate 
\begin{equation}\label{Eq:LogMomentIID}
\Lambda_S(\theta) = \log \left( \sum_i p_i e^{-\theta r_i} \right).
\end{equation}
Note that if two or more leafs of the decision tree correspond to the same leaf in the decision graph (and therefore result to the same transmission rate), it is equivalent to count them separately or as one term (with the cumulative probability as multiplication factor) in the above sum. In the following, unless otherwise noted, we will consider the decision tree (not the decision graph), in order to calculate the $p_i$ and $r_i$ in (\ref{Eq:LogMomentIID}). Also note that due to the shape of the function $\Lambda_A(\theta) - \Lambda_S(-\theta)$ (see Fig$.$~\ref{Fig:Figure_1}), when comparing two different functions it will be $\theta^*_{(1)}<\theta^*_{(2)}$ if and only if $\Lambda_A^{(1)}(\theta) - \Lambda_S^{(1)}(-\theta) > \Lambda_A^{(2)}(\theta) - \Lambda_S^{(2)}(-\theta), \forall \theta \in [\theta^*_{(1)}, \theta^*_{(2)}]$, and consequently when comparing two policies for the same arrival process we will have 
\begin{align}\label{Eq:OptEquivalence}
\theta^*(\mathbf{D}_1)&<\theta^*(\mathbf{D}_2) \Leftrightarrow \\
\nonumber \Lambda_S(-\theta, \mathbf{D}_1) &< \Lambda_S(-\theta, \mathbf{D}_2), \forall \theta \in [\theta^*(\mathbf{D}_1), \theta^*(\mathbf{D}_2)].
\end{align}

Coming to the format of the policy matrix $\mathbf{D}$, we will prove that if an element of $\mathbf{D}$ is $1$ then all elements below this element in the same column and all elements to the right of this element (up to the matrix diagonal) are also equal to $1$. We also note that all elements in the first column are equal to $0$ since if zero idle channels have been discovered, the SU should always continue sensing. The proof that the optimal policy matrix must have the described format is broken into the following two theorems.  

\begin{theorem} \label{th:alg1} \rm{
For any $k=0,1,\ldots,W-1$ and $i=0,1,\ldots,k-1$, if $\mathbf{D}_{k,i}=1$, then $\forall$~$j$ for which it holds $i<j\leq k$ it will be $\mathbf{D}_{k,j}=1$.}
\end{theorem}

\begin{IEEEproof}
For the optimum policy $\mathbf{D}$ (for which $\mathbf{D}_{k,i}=1$), we arrive at state $(k,i)$ with a certain probability (denoted by $p$) and we know that the decision maximizing $\theta^*$ is $\mathbf{D}_{k,i}=1$. Note that $p$ is the cumulative probability of all nodes in the decision tree corresponding to this particular node in the decision graph. Under the decision $\mathbf{D}_{k,i}=1$, this branch of the decision graph arrives to a leaf (no more decisions) and the resulting service rate is $i(K-k)c$. Therefore, the log moment generating function can be written as:  
\begin{equation}\label{Eq:Log_MGF_1}
\Lambda_S(-\theta, \mathbf{D})=\log\left(pe^{-\theta i(K-k)c}+\sum_{n=1}^Np_ne^{-\theta r_n}\right)
\end{equation}
where $\sum_{n=1}^Np_n=1-p$. 
Now consider any relaxed policy $\mathbf{G}$ derived by the optimal policy when we flip $\mathbf{D}_{k,i}$ from $1$ to $0$ and for any combination of decisions in the decision tree branch rooted at state $(k,i)$. For this relaxed policy, we have 
\begin{equation}\label{Eq:Log_MGF_2}
\Lambda_S(-\theta, \mathbf{G})=\log\left(p\sum_{m=1}^Mq_m e^{-\theta r_m}+\sum_{n=1}^Np_n e^{-\theta r_n}\right),
\end{equation}
and since by assumption $\mathbf{D}$ is optimal (no relaxed policy can improve over the optimum policy) and making use of (\ref{Eq:OptEquivalence})
$\Lambda_S(-\theta, \mathbf{G})>\Lambda_S(-\theta, \mathbf{D})$  $\forall \theta \in [\theta^*(\mathbf{G}), \theta^*(\mathbf{D})] $, results in 
\begin{equation}\label{Eq:Inequality_1}
\sum_{m=1}^Mq_m e^{-\theta r_m}>e^{-\theta i(K-k)c}\,\, \forall \theta \in [\theta^*(\mathbf{G}), \theta^*(\mathbf{D})].
\end{equation}
The potentially achievable rates $r_m$ in the new branch, if the relaxed policy $\mathbf{G}$ is used are:
\begin{align} 
\nonumber & i(K-k-1)c, (i+1)(K-k-1)c   \\
\nonumber & i(K-k-2)c, (i+1)(K-k-2)c, (i+2)(K-k-2)c,   \\
\nonumber & i(K-k-3)c, (i+1)(K-k-3)c, (i+2)(K-k-3)c, \\
          & (i+3)(K-k-3)c   \\ 
\nonumber & \hspace{0.5cm} \vdots 
\end{align}
Now for a fixed $j$ such that $i<j\leq k$, consider the policy $\mathbf{F}$ which is derived by $\mathbf{D}$ when setting $\mathbf{D}_{k,j}=0$ and the policy $\mathbf{H}$ derived by setting $\mathbf{D}_{k,j}=1$. Clearly, one of the two policies $\mathbf{F}$ or $\mathbf{H}$ is identical to $\mathbf{D}$. 
In both policies we reach the graph state $(k,j)$ with probability $\pi$ and so for $\mathbf{H}$ it holds:
\begin{equation}\label{Eq:Log_MGF_3}
\Lambda_S(-\theta, \mathbf{H})=\log\left(\pi e^{-\theta j(K-k)c}+\sum_{\ell=1}^L \pi_\ell e^{-\theta r_\ell}\right),
\end{equation}
and for $\mathbf{F}$:
\begin{equation}\label{Eq:Log_MGF_4}
\Lambda_S(-\theta, \mathbf{F})=\log\left(\pi\sum_{s=1}^S w_s e^{-\theta r_s}+\sum_{\ell=1}^L \pi_\ell e^{-\theta r_\ell}\right).
\end{equation}

The potentially achievable rates $r_s$ in the branch rooted at the state $(k,j)$ under policy $\mathbf{F}$ are 
\begin{align} 
\nonumber & j(K-k-1)c, (j+1)(K-k-1)c   \\
\nonumber & j(K-k-2)c, (j+1)(K-k-2)c, (j+2)(K-k-2)c,   \\
\nonumber & j(K-k-3)c, (j+1)(K-k-3)c, (j+2)(K-k-3)c, \\
          & (j+3)(K-k-3)c   \\ 
\nonumber & \hspace{0.5cm} \vdots 
\end{align}
So there is a one-to-one correspondence $r_m \leftrightarrow r_s$ and $M=S$. On the contrary, it is not necessary that $q_m = w_s$. However, for any policy $\mathbf{E}$ with $\mathbf{E}_{k,j}=0$ there is a matching relaxed policy $\mathbf{\Xi}$ with $\mathbf{\Xi}_{k,i}=0$ for which the tree starting at the graph state $(k,i)$ is the same as the tree starting at the graph state $(k,j)$ of $\mathbf{E}$. Therefore, for $\mathbf{\Phi}$ the matching relaxed policy of $\mathbf{F}$ it holds $q_m = w_s$ $\forall m=s$. Since (\ref{Eq:Inequality_1}) holds for any relaxed policy derived by $\mathbf{D}$ by setting $\mathbf{D}_{k,i}=0$ and for any combination of decisions in the branch following the state $(k,i)$, it also holds for the relaxed policy $\mathbf{\Phi}$.

For this policy we have that $\forall \theta \in [\theta^*(\mathbf{\Phi}), \theta^*(\mathbf{D})]$ holds:
\begin{align}
\nonumber & e^{-\theta i(K-k)c} < \sum_{m=1}^M q_m e^{-\theta r_m} \\
\nonumber \Rightarrow & e^{-\theta i(K-k)c} e^{-\theta (j-i)(K-k)c} \\
\label{Eq:CDF_gamma_d_alg1}  < & \sum_{m=1}^M q_m e^{-\theta \left(r_m + (j-i)(K-k)c\right)} \\
\nonumber \Rightarrow & e^{-\theta j(K-k)c} < \sum_{s=1}^S w_s e^{-\theta \left(r_m + (j-i)(K-k)c\right)}.
\end{align}
Now it can be easily verified that for any pair of $r_m$ and $r_s$ it holds $r_s = r_m + (j-i)(K-k)c$ and therefore
\begin{align}
\label{Eq:Inequality_2} & e^{-\theta j(K-k)c} < \sum_{s=1}^S w_s e^{-\theta r_s} \Rightarrow 
\\& \Lambda_S(-\theta, \mathbf{H})>\Lambda_S (-\theta, \mathbf{F}) 
\nonumber \forall \theta \in [\theta^*(\mathbf{\Phi}), \theta^*(\mathbf{D})].
\end{align}
\begin{figure}[t!] 
\centering
\includegraphics[width=3.2in]{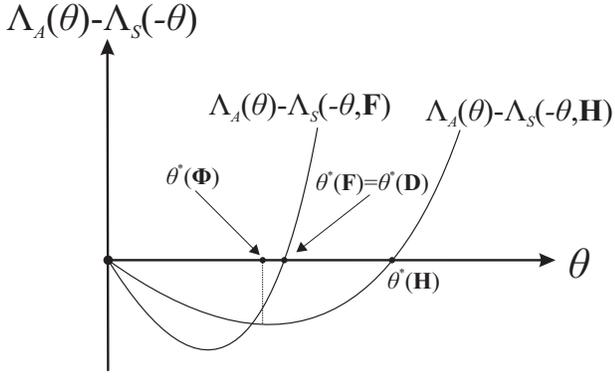} 
\caption{Graphical justification of $\theta^*(\mathbf{H})>\theta^*(\mathbf{F})$.}
\label{Fig:Figure_3}
\end{figure}
Recall that exactly one of $\mathbf{H}$ and $\mathbf{F}$ is identical to $\mathbf{D}$. Assume $\mathbf{F} = \mathbf{D}$. Then $\theta^*(\mathbf{F}) = \theta^*(\mathbf{D})$ and the function $\Lambda_A(\theta) - \Lambda_S(-\theta, \mathbf{H}) < \Lambda_A(\theta) - \Lambda_S(-\theta, \mathbf{F})$ in a non-empty interval $[\theta^*(\mathbf{\Phi}), \theta^*(\mathbf{F})]$. Therefore, the root $\theta^*(\mathbf{H})$  of $\Lambda_A(\theta) - \Lambda_S(-\theta, \mathbf{H})$ is larger than the root $\theta^*(\mathbf{F})$ of $\Lambda_A(\theta) - \Lambda_S(-\theta, \mathbf{F})$ (see Fig$.$~\ref{Fig:Figure_3}) and this contradicts our hypothesis. Hence, it must be $\mathbf{H} = \mathbf{D}$. 
\end{IEEEproof} \vspace{1mm}

\begin{theorem} \label{th:alg2} \rm{
For any $k=0,1,\ldots,W-2$ and $i=0,1,\ldots,k$, if $\mathbf{D}_{k,i}=1$, then $\forall$~$\ell$ for which it holds $k<\ell\leq W-1$ it will be $\mathbf{D}_{\ell,i}=1$.}
\end{theorem}

\begin{IEEEproof}
As the assumptions in this theorem are the same as in Theorem \ref{th:alg1}, the starting point of the proof is the same, i$.$e$.$, policies $\mathbf{D}$ and $\mathbf{G}$ for which (\ref{Eq:Log_MGF_1}), (\ref{Eq:Log_MGF_2}), and (\ref{Eq:Inequality_1}) hold.  

Now for a fixed $\ell$ such that $k<\ell\leq W-1$, consider the policy $\mathbf{F}$ which is derived by the optimal $\mathbf{D}$ by setting $\mathbf{D}_{\ell,i}=0$, and the policy $\mathbf{H}$ derived by setting $\mathbf{D}_{\ell,i}=1$. (Note that in this proof we redefine previously defined symbols with a different meaning). Obviously, one of the two policies $\mathbf{F}$ or $\mathbf{H}$ is identical to $\mathbf{D}$.  

In both policies we reach the state $(\ell,i)$ with probability $\pi$ and so for $\mathbf{H}$ it holds:
\begin{equation}\label{Eq:Log_MGF_5}
\Lambda_S(-\theta, \mathbf{H})=\log\left(\pi e^{-\theta i(K-\ell)c}+\sum_{j=1}^J \pi_j e^{-\theta r_j}\right),
\end{equation}
and for $\mathbf{F}$:
\begin{equation}\label{Eq:Log_MGF_6}
\Lambda_S(-\theta, \mathbf{F})=\log\left(\pi\sum_{s=1}^S w_s e^{-\theta r_s}+\sum_{j=1}^J \pi_\ell e^{-\theta r_j}\right).
\end{equation}
For the policy $\mathbf{F}$ we can construct a matching relaxed policy $\mathbf{\Phi}$ such that the decision branch rooted at the state $(k,i)$ of $\mathbf{\Phi}$ is the same as the decision tree of $\mathbf{F}$ rooted at the state $(\ell,i)$. The probabilities of reaching equivalent states at the two branches are the same whereas, the rates $r_s$ at the considered branch in $\mathbf{F}$ are related to the rates $r_m$ at the matching branch in $\mathbf{\Phi}$ according to $r_s = r_m - (\ell - k) d_s c$, where $d_s$ can take integer values greater than or equal to $i$. So, there is a one-to-one correspondence $r_s \leftrightarrow r_m$ and $S=M$. (Note that as the remaining sensing slots at state $(\ell,i)$ are fewer than the remaining sensing slots at the state $(k,i)$, in the considered branch of the matching policy all sensing will stop at most at the $K+k-\ell$ sensing slot). 
\renewcommand\thealgorithm{A}
\begin{algorithm}[!t] \caption{} \label{alg:alg1}
\smallskip
\begin{algorithmic}
\STATE \hspace{-4mm} \texttt{Initialization:} \\
  \hspace{0.1cm}Construct a lower diagonal matrix $\mathbf{F}$ such that:\\
  \hspace{0.1cm}$\mathbf{F}_{k,0}=0$, $\forall$ $k=0,1,\ldots,W-1$, and \\ 
  \hspace{0.1cm}$\mathbf{F}_{k,s}=1$, $\forall$ $k=0,1,\ldots,W-1$ and~$1\leq s \leq k$. \\
  \hspace{0.1cm}Exhaustively find all combinations of $W-1$ numbers \\
	\hspace{0.1cm}$x_1, x_2, \dots, x_{W-1}$ s.t.
$1 \leq x_{W-1} \leq x_{W-2} \leq \dots \leq x_1$. 
\STATE \hspace{-4mm} \texttt{Main body:} \\
\hspace{0.1cm}\texttt{for} each one of the above combinations\\ 
\hspace{0.25cm} Construct a policy matrix $\mathbf{D}$ as follows:\\
\hspace{0.25cm} Start with matrix $\mathbf{F}$. For all its rows turn the first $x_j$ \\
\hspace{0.25cm} elements of row $j$ to $0$ to get $\mathbf{D}$. Compute $\theta^*(\mathbf{D})$.\\
\hspace{0.25cm} Keep the largest $\theta^*(\mathbf{D})$ over all iterations.  \\
\hspace{0.1mm} \texttt{endfor} 
\STATE \hspace{-4mm} \texttt{Output:} The largest $\theta^*(\mathbf{D})$ and associated  $\mathbf{D}$ 
\end{algorithmic}
\end{algorithm} \vspace{-1mm}

For the relaxed policy $\mathbf{\Phi}$, (\ref{Eq:Inequality_1}) holds, since (\ref{Eq:Inequality_1}) holds for any relaxed policy derived by $\mathbf{D}$ by setting $\mathbf{D}_{k,i}=0$, and for any combination of decisions in the branch following the state $(k,i)$.

Therefore, $\forall \theta \in [\theta^*(\mathbf{\Phi}), \theta^*(\mathbf{D})]$ holds:
\begin{align}
\nonumber & e^{-\theta i(K-k)c} < \sum_{m=1}^M q_m e^{-\theta r_m} \\
\nonumber \Rightarrow & e^{-\theta i(K-k)c} e^{-\theta i(k-\ell)c} < e^{-\theta i(k-\ell)c} \sum_{m=1}^M q_m e^{-\theta r_m} \\
\nonumber \Rightarrow & e^{-\theta i(K-\ell)c} < e^{-\theta i(k-\ell)c} \sum_{s=1}^S w_s e^{-\theta \left[ r_s + (\ell-k)d_s c\right]} \\
\label{Eq:CDF_gamma_d_alg2}  = & \sum_{s=1}^S w_s e^{-\theta \left[ i(k-\ell)c + r_s + (\ell-k)d_s c\right]} \\
\nonumber = & \sum_{s=1}^S w_s e^{-\theta r_s} e^{-\theta (\ell-k)(d_s-i) c}  <  \sum_{s=1}^S w_s e^{-\theta r_s} \\
\nonumber \Rightarrow & \Lambda_S(-\theta, \mathbf{H})>\Lambda_S (-\theta, \mathbf{F}).
\end{align}
Similarly to the proof of Theorem \ref{th:alg1}, exactly one of $\mathbf{H}$ and $\mathbf{F}$ is equal to $\mathbf{D}$. Assume $\mathbf{F} = \mathbf{D}$. Then $\theta^*(\mathbf{F}) = \theta^*(\mathbf{D})$ and the function $\Lambda_A(\theta) - \Lambda_S(-\theta, \mathbf{H}) < \Lambda_A(\theta) - \Lambda_S(-\theta, \mathbf{F})$ in a non-empty interval $[\theta^*(\mathbf{\Phi}), \theta^*(\mathbf{F})]$. Therefore, the root $\theta^*(\mathbf{H})$  of $\Lambda_A(\theta) - \Lambda_S(-\theta, \mathbf{H})$ is larger than the root $\theta^*(\mathbf{F})$ of $\Lambda_A(\theta) - \Lambda_S(-\theta, \mathbf{F})$, and this contradicts our hypothesis. Hence, it must be $\mathbf{H} = \mathbf{D}$. 
\end{IEEEproof} \vspace{1mm}

An obvious algorithm that can exploit the described format of the optimal policy matrix more efficiently than exhaustive search is shown in Algorithm~\ref{alg:alg1}. An even more efficient heuristic algorithm is shown in Algorithm~\ref{alg:alg2}. Algorithm~\ref{alg:alg2} is justified by observing that when we flip bits down a column, flipping the next bit from $1$ to $0$ cannot make us reconsider previous decisions to flip in the same column (as this would create a $1$ above a $0$). Also, if a bit is not flipped, there is no point to consider subsequent bits in the same column, so we can move to the next column. Finally, if there exists a bit to the left of the considered bit that is equal to $1$, the considered bit cannot be changed to $0$, and we can safely move to the next column. Although Algorithm~\ref{alg:alg2} has been tested and found to discover the same optimal policy as Algorithm~\ref{alg:alg1} in all tests, its optimality has not been formally proven due to the following consideration: when flipping a bit from $1$ to $0$ in a column, can this make us reconsider an already flipped bit in a previously considered column and in a row below the one of the currently flipped bit? This remains an open question.   

\renewcommand\thealgorithm{B}
\begin{algorithm}[!t] \caption{} \label{alg:alg2}
\smallskip
\begin{algorithmic}
\STATE \hspace{-4mm} \texttt{Initialization:} \\
\hspace{0.1cm}Lower diagonal policy matrix $\mathbf{D}$ such that:\\
\hspace{0.1cm}$\mathbf{D}_{k,0}=0$, $\forall$ $k=0,1,\ldots,W-1$, and \\ 
\hspace{0.1cm}$\mathbf{D}_{k,s}=1$, $\forall$ $k=0,1,\ldots,W-1$ and~$1\leq s \leq k$. \\
\hspace{0.1cm}$\theta^*_{opt} = \theta^*(\mathbf{D})$. 
\STATE \hspace{-4mm} \texttt{Main body:} \\
\hspace{0.1cm}\texttt{for} $j=1,2,\ldots,W-1$\\
\hspace{0.6cm}\texttt{for} $i=j,j+1,\ldots,W-1$\\
\hspace{1.1cm}\texttt{if} $\mathbf{D}_{i,j}=1$ \\
\hspace{1.6cm}\texttt{if} $\exists k<j \ s.t. \ \mathbf{D}_{i,k}=1$\\
\hspace{2.1cm}\texttt{break};\\
\hspace{1.6cm}\texttt{endif}\\
\hspace{1.6cm}$\mathbf{D}_{i,j}=0$;\\
\hspace{1.6cm}\texttt{if} $\theta^*(\mathbf{D})>\theta^*_{opt}$ \\			
\hspace{2.1cm}$\theta^*_{opt}=\theta^*(\mathbf{D})$;\\
\hspace{1.6cm}\texttt{else}\\
\hspace{2.1cm}$\mathbf{D}_{i,j}=1$;\\
\hspace{2.1cm}\texttt{break};\\
\hspace{1.6cm}\texttt{endif}\\
\hspace{1.1cm}\texttt{endif}\\
\hspace{0.6cm}\texttt{endfor}\\
\hspace{0.1cm}\texttt{endfor}
\STATE \hspace{-4mm}\texttt{Output:} The largest $\theta^*(\mathbf{D})$ and associated  $\mathbf{D}$ 
\end{algorithmic}
\end{algorithm} \vspace{-1mm}

\section{Results and Discussion}\label{sec:Results} 
\begin{figure}[t!] 
\centering
\includegraphics[width=3.2in]{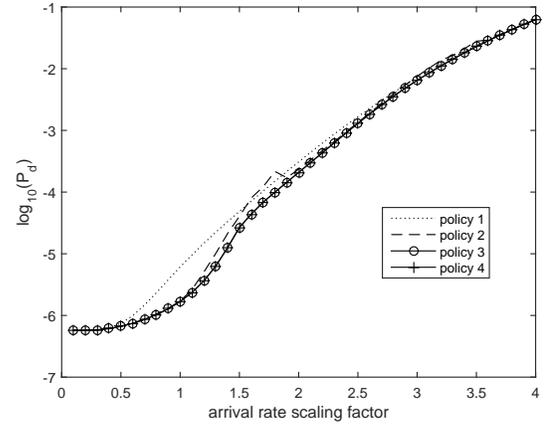} 
\caption{Comparing the four policies for optimizing the probability of excessive delay.}
\label{Fig:Figure_4}
\end{figure}

In this Section we present preliminary results in order to demonstrate the proposed algorithms and derived policies. For simplicity, we model the arrival process as a Markov chain with 5 states. In Fig$.$ \ref{Fig:Figure_4}, we are showing the LD approximations for the excessive delay probabilities under four policies as a function of a scaling multiplicative factor to all arrival rates. Policy $1$ is the average throughput maximizing policy. Policy $2$ is a DP policy which uses the LD exponent as the cost function which tries to optimize. Policy $3$ is derived by running Algorithm~\ref{alg:alg1} and policy $4$ by using Algorithm~\ref{alg:alg2}. In this example $K=W=10$, $c=1$, $p_{idle}=0.55$ and $D_{max}=2$. It is obvious that policies $3$ and $4$ coincide and are optimal, while the other two policies are suboptimal. The optimality gap of the suboptimal policies is a function of the problem parameters. In many cases, especially for $D_{max}>1$, policy $1$ is a good approximation to the optimum policy.  

\section{Conclusion}\label{sec:Conclusions}
Optimizing the probability of excessive delay in a queue is different than maximizing the average service rate.
In this paper, based on LD approximations for the probability of large delays, we proposed two efficient algorithms for deriving optimal policies for the sensing / transmitting trade-off in hardware constrained DSA. One of these two policies was proven to be optimal, while the other is more efficient and seems to provide the optimal policy in all cases. We compared these policies to the maximizing average throughput policy and a DP derived policy, and concluded that these two later policies are indeed suboptimal in various regions of the parameter space.  


\bibliographystyle{IEEEtran}
\bibliography{IEEEabrv,references}
\end{document}